\def\g2{{$(g-2)$} }

\documentclass[
    ,final            
  ]
  {aipproc}

\layoutstyle{6x9}

\begin{document}

\title{Lepton Dipole Moments
}

\author{B. Lee Roberts}
{
  address={roberts@bu.edu\\
Department of Physics\\
Boston University\\
Boston, MA 02215 USA 
}
}

\begin{abstract}
From the famous experiments of Stern and Gerlach to the present,
measurements of magnetic dipole moments, and searches for electric
dipole moments of ``elementary'' particles have played a major
role in our understanding of sub-atomic physics.  In this talk
I discuss the progress on measurements and theory of the 
magnetic dipole moments of the electron and muon.  I also discuss
a new proposal to search for a permanent electric dipole moment (EDM) of
the muon and put it into the more general context of other EDM searches.
\end{abstract}

\maketitle


\section{Introduction and theory of the lepton anomalies}

Over the past 82 years, the study of dipole moments of 
elementary particles has provided a wealth of information on 
subatomic physics.  From the pioneering work of Stern\cite{stern}
through the discovery of the large anomalous magnetic moments
of the proton\cite{sternp} and neutron\cite{nmdm}, the ground
work was laid for the discovery of spin, of radiative corrections
and the renormalizable theory of QED, of the quark structure of
baryons and the development of QCD.  

A charged particle with spin $\vec s$ has a magnetic moment
\begin{equation}
 \vec \mu_s = g_s ( {e \over 2m} ) \vec s;
\qquad \mu = (1 + a){e \hbar \over 2m}; \qquad a \equiv { (g_s -2) \over 2};
\end{equation}
where $g_s$ is called the gyromagnetic ratio.  The expression in
the middle is the quantity one finds listed in the Particle
Data Tables.\cite{pdg}
The quantity $a$ is the anomalous magnetic dipole moment (or simply
the anomaly) which is related to the $g$-factor in the right-hand equation.

\begin{figure}[h!]
  \includegraphics[height=.10\textheight]{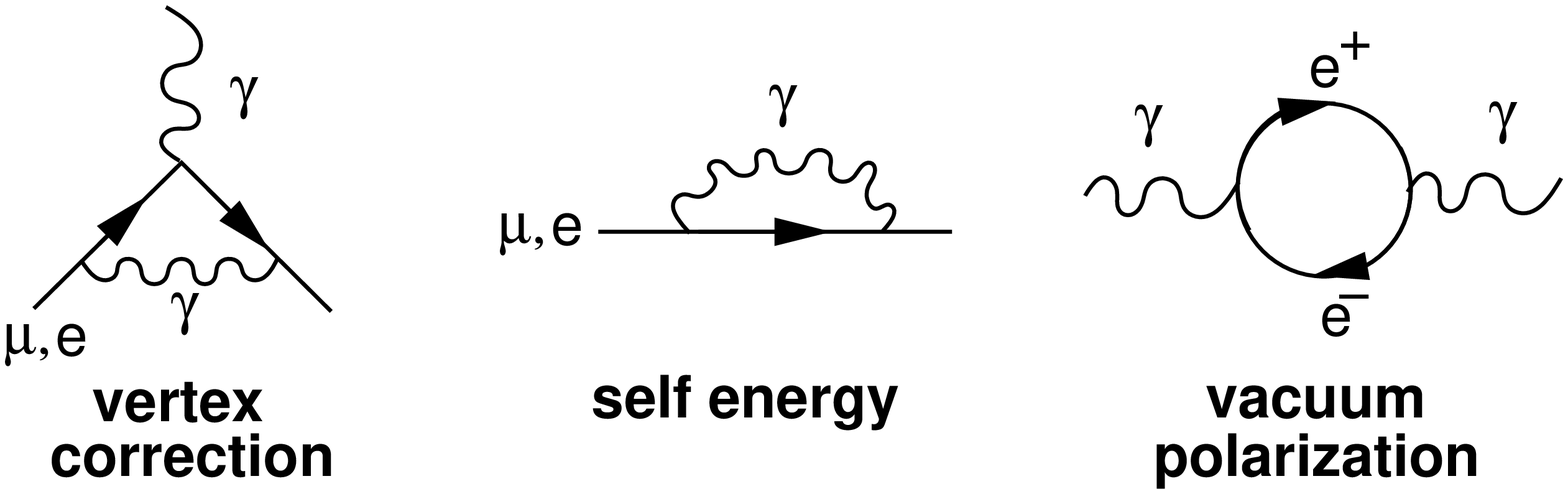}
  \caption{Three examples of radiative corrections.  The middle
term is absorbed into the dressed mass, but the vertex correction and
vacuum polarization play an important role in the anomaly.}  
\label{fg:radcor}
\end{figure}

The Dirac equation tells us that $g\equiv 2 $ for spin angular momentum,
and is unity for orbital angular momentum (the
latter having been verified experimentally\cite{kusch}).  
This can be seen from the
non-relativistic reduction for an electron in a weak magnetic field:
\begin{equation}
i \hbar {\partial \psi \over \partial t} =
\left[ {p^2 \over 2m} - {e \over 2 m } (\vec L + 
{2}  {\vec S}) {\cdot \vec B} \right] \psi ,
\end{equation}
and the subscript on $g$ is dropped in the following discussion.
For point particles, the anomaly arises from radiative corrections,
three examples of which are shown in Fig.~\ref{fg:radcor}. The 
``vertex'' correction and vacuum polarization are important in the
lepton anomaly, while the self-energy term is included in the 
dressed mass.  The situation for baryons is quite different, since their
internal quark structure gives them large anomalies.

The vertex correction in lowest-order gives the famous 
Schwinger\cite{schwinger}
result, \break $a =\alpha/2 \pi$,
which was verified experimentally
by Foley and Kusch.\cite{kusch}
To lowest order,
 the S-matrix element\cite{bd}
for a charged particle in a magnetic field is given by 

\begin{equation}
- i e \bar u(p')
\left[ {(p + p')_{\lambda} \over 2 m} +
{
\left( 1 + {\alpha \over 2 \pi}\right)
{
{i \sigma_{\lambda \nu} q^{\nu}} \over 2 m
}}
\right]
 u(p) A^{\lambda}(q)
\end{equation}

In general {$a$} (or $g$) is an expansion 
in $ \left({\alpha \over  \pi}\right)$,
\begin{equation}
a = C_1\ \left( {\alpha \over  \pi}\right) 
+ C_2\  \left( {\alpha \over  \pi}\right)^2
+C_3\ \left( {\alpha \over  \pi}\right)^3 
+ C_4\ \left( {\alpha \over  \pi}\right)^4 + \cdots
\end{equation}
with 1 diagram for the Schwinger (second-order) contribution,
5 for the fourth order, 40 for the sixth order, 891 for the eighth
order.

The QED contributions to electron and muon \g2 have now been calculated
through eighth order, $(\alpha/\pi)^4$ and the the 
tenth-order contribution has been estimated.\cite{kinqed}  The first few
orders are shown schematically below in Fig. \ref{fg:aexpan}.

\begin{figure}[h!]
  \includegraphics[height=.08\textheight]{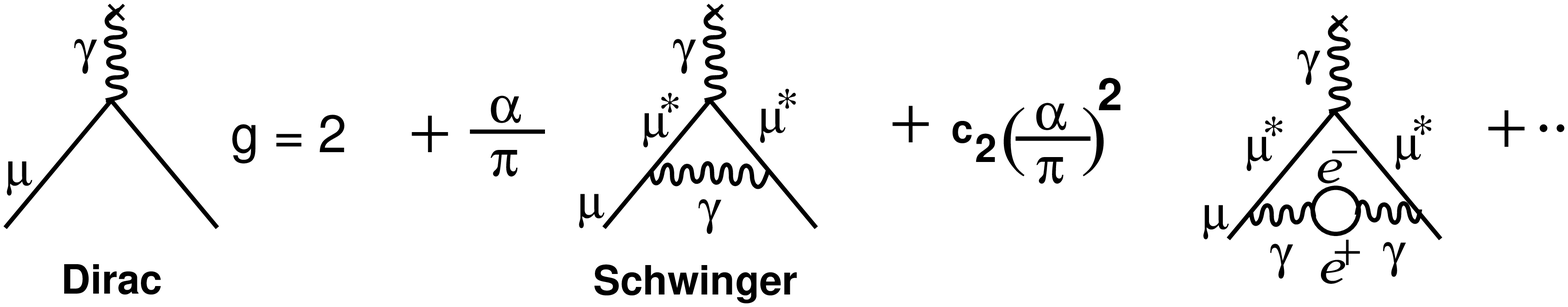}
  \caption{A schematic of the first few terms in the QED expansion
for the muon. The vacuum polarization term shown is one of
five of order $(\alpha/\pi)^2$.}
\label{fg:aexpan}
\end{figure}

While magnetic dipole 
moments (MDMs) are a natural property of charged particles with spin,
electric dipole moments (EDMs) are forbidden both by parity and 
by time reversal symmetry. Interestingly enough,
Purcell and Ramsey\cite{pr} suggested in 1950 that a measurement of 
the neutron EDM would be a good way to search for parity violation,
well in advance of the paper by Lee and Yang.
After the discovery of parity violation,
Landau\cite{landau} and Ramsey\cite{ramsey}  
pointed out that an EDM would violate both
{\sl P} and {\sl T} symmetries. This can be seen by examining
the Hamiltonian for a spin one-half particle in the presence of
both an electric and magnetic field,
${\mathcal H} = - \vec \mu \cdot \vec B  - \vec d \cdot \vec E$.
The transformation properties of $\vec E$, $\vec B$, $\vec \mu$ and $\vec d$
are given in the Table \ref{tb:tranprop}, and we see that while
$\vec \mu \cdot \vec B$ is even under all three,
$\vec d \cdot \vec E$ is odd under both {\sl P} and
{\sl T}.  Thus
the existence of an EDM implies that both {\sl P} and {\sl T} are violated.
In the context of {\sl CPT} symmetry, an EDM implies {\sl CP} violation.
The standard model value for the
electron and muon EDM is $\leq 10^{-35}$ $e$-cm, well beyond the reach of 
experiments (which are at the $10^{-26}$ $e$-cm level). 
Observation of a non-zero $e$ or $\mu$ EDM
would be a clear signal for
new physics.

\begin{table}[htb]
\begin{tabular}{cccc} \hline
      & \tablehead{1}{c}{b}{$\vec E$ }
      & \tablehead{1}{c}{b}{$\vec B$ }
      & \tablehead{1}{c}{b}{$\vec \mu$ or  $\vec d$} \\
\hline
{\sl P} & - & + & + \\
{\sl C} & - & - & - \\
{\sl T} & + & - & - \\
\hline
\end{tabular}
\caption{Transformation properties of the magnetic and electric fields and
dipole moments.}
\label{tb:tranprop}
\end{table}

The connection between the magnetic and electric dipole moments can
be seen by writing the interaction Lagrangian as

\begin{equation}
{\mathcal{L}}_{dm} = {1\over 2} \left[ D\bar \mu \sigma^{\alpha \beta}
{1 + \gamma_5 \over 2} \  + \ D^* \bar \mu \sigma^{\alpha \beta}
{1 - \gamma_5 \over 2}\right] \mu F_{\alpha \beta}
\end{equation}
where the dipole operator $D$ has 
${\rm Re}\ D\  =\  a_{\mu} {e \over 2 m_{\mu}}$  and
${\rm Im}\ D\  =\  d_{\mu}$.

The standard model value of $a$ has three contributions from radiative
processes: QED loops containing leptons ($e,\mu,\tau$) and photons; 
hadronic loops containing hadrons in vacuum polarization loops; 
and weak loops involving the weak gauge bosons $W,Z,$ and Higgs. Thus
$a_{e,\mu}{\rm ( SM)}
= a_{e,\mu}({\rm QED}) + a_{e,\mu}({\rm hadronic}) +
a_{e,\mu}({\rm weak})$.
A difference between the experimental value and the standard model
prediction would signify the presence of new physics beyond the
standard model.  Examples of such potential contributions are
lepton substructure, extra gauge bosons, anomalous $W-\gamma$
couplings, or the existence of supersymmetric partners of the
leptons and gauge bosons.

The electron anomaly is now measured to a relative precision
of about four parts in a 
billion (ppb),\cite{eg2} and the muon is measured to 0.7 parts per 
million (ppm).\cite{bennett}  The relative contributions of 
heavier particles to $a$ scales as $(m_e/m_{\mu})^2$, and the muon
has a sensitivity factor of about 40,000 over the electron 
to higher mass scale radiative corrections. This gives 
the muon an overall
advantage of 230 in measurable sensitivity to larger mass scales,
including new physics.
At a precision of 0.7 ppm, the muon anomaly
is sensitive to $\geq 100$ GeV scale physics.

In fact,  the contribution of anything heavier than an electron to
the electron anomaly is at the level
of about 3 ppb. So while the the electron $(g-2)$ experiments are
triumphs of experimental and theoretical physics, they are purely a test
of QED. Since the independent measurements of $\alpha$ 
are less precise (7.4 ppb)
than the present accuracy on the electron anomaly,
the measurement of $a_e$ has been used to
determine the best measurement of the fine-structure 
constant.\cite{kinalpha}  The uncertainty in $\alpha$ is not an issue
for $a_{\mu}$.

\begin{figure}[h!]
  \includegraphics[height=.12\textheight]{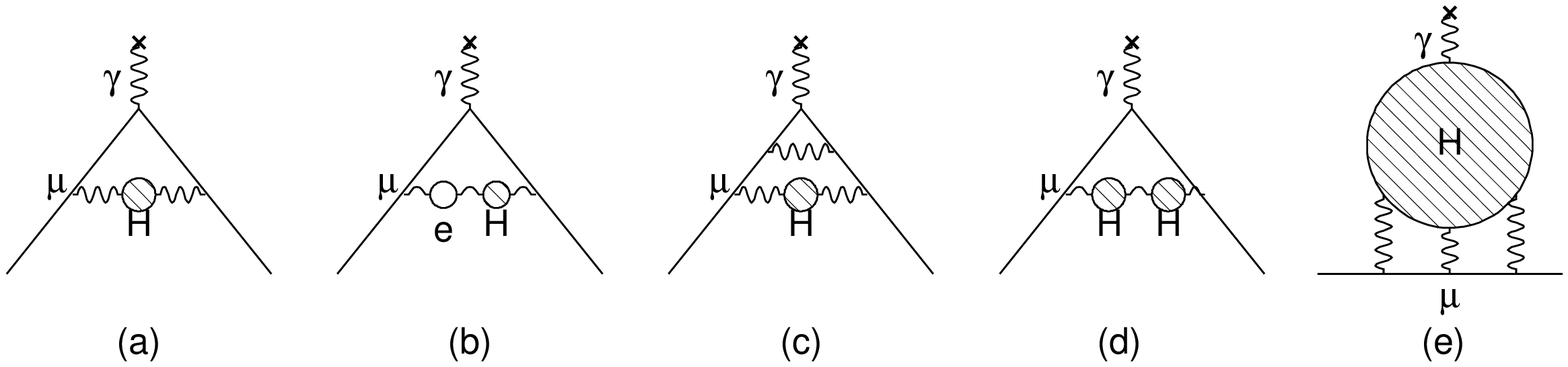}
  \caption{The hadronic contribution to the muon anomaly, where the
dominant contribution comes from (a).  The hadronic light-by-light
contribution is shown in (e).}
\label{fg:had}
\end{figure}

The CERN experiment\cite{cern3}
observed the contribution of hadronic vacuum polarization 
shown in Fig. \ref{fg:had}(a)  at the
10 standard deviation level.  Unfortunately, the hadronic contribution
cannot be calculated directly from QCD, since the energy scale
is very low ($m_{\mu} c^2$), although Blum\cite{blum} has performed 
a proof of principle calculation on the lattice.  
Fortunately dispersion theory 
gives a relationship between the vacuum polarization loop
and the cross section for $e^+ e^- \rightarrow {\rm hadrons}$,
\begin{equation}
a_{\mu}({\rm Had;1})=({\alpha m_{\mu}\over 3\pi})^2
\int^{\infty} _{4m_{\pi}^2} {ds \over s^2}K(s)R(s), \quad
{\rm where} \quad
R\equiv{ {\sigma_{\rm tot}(e^+e^-\to{\rm hadrons})} \over
\sigma_{\rm tot}(e^+e^-\to\mu^+\mu^-)}
\end{equation}
and experimental
data are used as input.
The factor $s^{-2}$ in the dispersion relation, means that values
of  $R(s)$ at low energies (the $\rho$ resonance) dominate the determination of
$a_{\mu}({\rm Had;1})$. This information can also be obtained
from hadronic $\tau^-$ decays such as
$\tau^- \rightarrow \pi^- \pi^0 \nu_{\tau} $, which can be 
related to $e^+e^-$ annihilation through the CVC hypothesis and
isospin conservation.\cite{dh98,dehz}

The quantity $R$ is not directly measured.
 The cross-section for $e^+e^-$ is
determined using some other normalization, followed by
careful subtractions for initial
state radiation, vacuum polarization etc.,
 and the denominator is calculated from QED.
Most of these effects would cancel in the ratio if $R(s)$ were measured 
directly.

Knowledge of the hadronic contribution has steadily improved over
the past 15 years.  When the muon $(g-2)$ experiment E821 began at Brookhaven
in the early 1980s, $a_{\mu}({\rm Had;1})$
was known to about 5 ppm.
Now the uncertainty is about 0.5 ppm.  This progress has not come without
some pain.  The hadronic light-by-light contribution has changed sign
twice, with the positive sign now having been confirmed by a number of 
authors.\cite{hlbl}  New high-precision data from Novosibirsk on
$e^+e^-\rightarrow {\rm hadrons}$ lowered both the value and the
error on $a_{\mu}({\rm Had;1})$.\cite{cmd0,dehz}  
The value obtained from $\tau$-decay
differed from that using  $e^+e^-$.\cite{dehz}  Recently a normalization error
was uncovered, which moves the value of $a_{\mu}({\rm Had;1})$
obtained from $e^+e^-$ closer to that obtained from 
$\tau$-decay,\cite{novo} and a revised value\cite{dehz2} 
of $a_{\mu}({\rm Had;1})$
shows a difference between the standard model value and
the experimental result from E821 
to be: $(22.1\pm 7.2 \pm 3.5 \pm 8.0)\times 10^{-10}$ ($1.9 \sigma$) 
and $(7.4 \pm 5.8 \pm 3.5 \pm 8.0)\times 10^{-10}$
($0.7 \sigma$) for the $e^+e^-$ and $\tau$-based estimates
respectively.  The second
error is from the hadronic light-by-light 
contribution and the third one from E821. Further measurements of the
hadronic cross section are underway at Frascati and BaBar, using 
initial state radiation to lower $\sqrt s$ below the beam energy
(sometimes called radiative return).

\begin{figure}[h!]
  \includegraphics[height=.10\textheight]{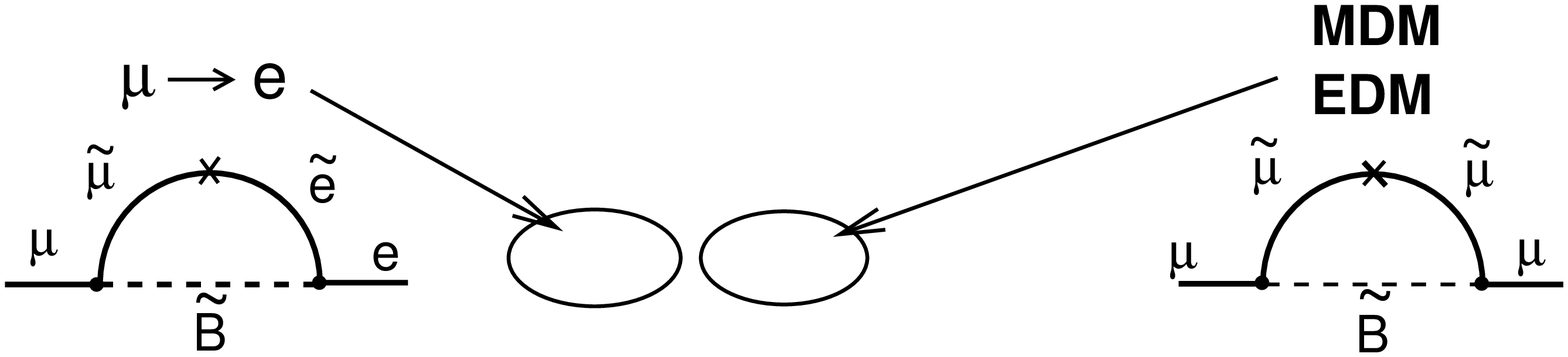}
  \caption{The supersymmetric contributions to the anomaly, and to
$\mu \rightarrow e$ conversion, showing the relevant slepton mixing matrix
elements. The MDM and EDM give the real and imaginary parts of
the matrix element, respectively.}  
\label{fg:susy}
\end{figure}

One of the very useful roles measurements of $a_{\mu}$ have played in the 
past is placing serious restrictions on physics beyond the standard model.
With the development of supersymmetric theories as a favored scheme of
physics beyond the standard model, interest in the experimental and
theoretical value of $a_{\mu}$ has grown substantially.  SUSY contributions
to $a_{\mu}$ could be at a measurable level in a broad range of models.
Furthermore, there is a complementarity between the SUSY contributions 
to the MDM, EDM and transition moment for $\mu \rightarrow e$.  
The MDM and EDM are related to the real and
imaginary parts of the diagonal element of the slepton 
mixing matrix, and the transition moment is related to the
off diagonal one, as shown in Fig. \ref{fg:susy}.

\section{Measurement of the muon anomaly}
The method used in the third
CERN experiment and the BNL experiment are
very similar, save the use of direct muon 
injection\cite{kick} into the storage ring,\cite{mag,inf}
which was developed by the E821 collaboration.  These
experiments are based on the
fact that for  $a_{\mu} > 0$ the spin 
precesses faster than
the momentum vector when a muon travels transversely to a 
magnetic field.  The spin precession frequency $\omega_S$
consists of the Larmor and Thomas spin-precession terms. The
spin frequency $\omega_S$, the momentum   
precession (cyclotron) frequency $\omega_C$, and the difference
frequency $\omega_a$ are given by 
\begin{equation}
 \omega_S = {geB \over 2 m c} + (1-\gamma) {e B \over \gamma mc};\qquad
 \omega_C = {e B \over mc \gamma}; \qquad
\omega_a = \omega_S - \omega_C = \left({g-2 \over 2}\right) {eB \over mc}.
\label{eq:omeganoE}
\end{equation}
The difference frequency is the frequency with which the spin
precesses relative to the momentum, which is  proportional to
the anomaly, rather than to $g$.
A precision measurement of $a_{\mu}$ requires precision measurements
of the precession frequency $\omega_a$  and the magnetic field,
which is expressed as the free-proton precession frequency
$\omega_p$ in the storage ring magnetic field.

The muon frequency can be measured as accurately as the counting
statistics and detector apparatus permit.  
The design goal for the NMR magnetometer and calibration system
was a field accuracy of 0.1 ppm.  The $B$ which enters in 
Eq. \ref{eq:omeganoE} is the average field seen by the ensemble of muons
in the storage ring.
The need for vertical focusing implies that a gradient field is needed,
but the usual magnetic gradient used in storage rings is ruled out,
since a sufficient magnetic gradient for vertical focusing would
spoil the ability to use NMR to 
measure the magnetic field to the necessary accuracy.
An electric quadrupole is used for vertical focusing, 
taking advantage of the 
``magic''~$\gamma=29.3$ at which an electric field does not contribute to
the spin motion relative to the momentum. With both an electric
and a magnetic field, the spin difference frequency is given by
\begin{equation}
\vec \omega_a = - {e \over mc}
\left[ a_{\mu} \vec B -
\left( a_{\mu}- {1 \over \gamma^2 - 1}\right) \vec \beta \times \vec E
\right],
\label{eq:tbmt}
\end{equation}
which reduces to Eq. \ref{eq:omeganoE} in the absence of an electric field.
For muons with $\gamma = 29.3$ in an electric field alone,
the spin would follow the momentum vector.

The experimental signal is the $e^{\pm}$ from $\mu^{\pm}$ decay, which 
were detected by lead-scintillating
fiber calorimeters.\cite{det}  The time and energy of each event was
 stored for analysis offline. 
Muon decay is a three-body decay, so the 3.1 GeV muons produce a continuum
of positrons (electrons) from the end-point energy down.  Since the highest
energy  $e^{\pm}$ are correlated with the muon spin, if one counts high-energy 
 $e^{\pm}$ as a function of time, one gets an exponential from muon decay
modulated by the $(g-2)$ precession. The expected form for the positron time
spectrum is $f(t) =  {N_0} e^{- \lambda t } 
[ 1 + {A} \cos ({\omega_a} t + {\phi})] $, however in analyzing the
data it is  necessary
to take a number of small effects into account in order to obtain
a satisfactory $\chi^2$ for the fit.\cite{bennett}

\begin{figure}[h!]
  \includegraphics[height=.25\textheight]{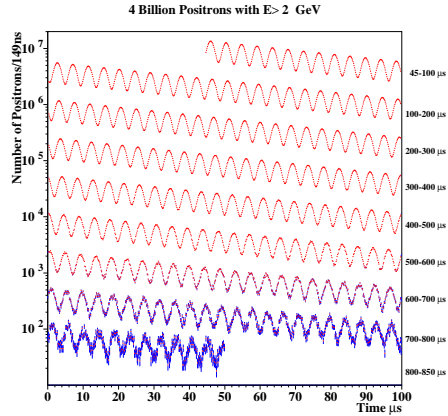}
  \caption{The time spectrum of positrons with energy greater than
2.0 GeV from the Y2000 run.  The endpoint energy is 3.1 GeV.
The time interval for each of the diagonal ``wiggles'' is given
on the right.}
\label{fg:wig00}
\end{figure}

\begin{figure}[h!]
  \includegraphics[height=.15\textheight]{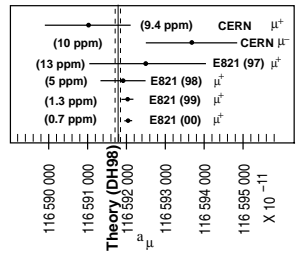}
  \caption{Measurements of $a_{\mu}$. The theory comes from
Davier and H\"ocker\cite{dh98} with the sign of the hadronic
light by light corrected. }
\label{fg:amu}
\end{figure}

The experimental results thus far are shown in Fig. \ref{fg:amu}, with
the average 
\begin{equation}
a_\mu(\rm{exp}) = 11\,659\,203(8) \times 10^{-10}
\qquad (\pm 0.7\ {\rm ppm})
\end{equation}
being dominated by results from E821. The theory value does
not reflect the re-analysis just made available,\cite{dehz2}
but rather $a_{\mu}({\rm Had;1})$ determined from
older data\cite{dh98} is shown. One additional
data set from E821 is being analyzed; the only data set obtained with
negative muons.  The result should be finalized in Fall '03 with
an expected uncertainty between 0.7 and 0.8 ppm. We are exploring the
possibility of an upgraded experiment at Brookhaven or at J-PARC\cite{g2jparc}
which could reach 0.1 ppm (BNL) or 0.06 ppm (J-PARC) precision.

\section{Measurement of the electron Anomaly}

While measurements of the electron anomaly have a history stretching
back to Kusch,\cite{kusch} a major breakthrough in precision came
with the trap experiments of Dehmelt.\cite{eg2}  Single electrons
were captured in a Penning trap, and the difference (beat) frequency
$\omega_a$ was measured (see Eq. \ref{eq:omeganoE}). Conventional
resonance techniques were employed to measure $\omega_c$, and 
this work was carried out using a hyperbolic
trap where the cavity shifts of the measured magnetic moments placed
serious limitations on the precision available.\cite{gtb}

\begin{figure}[h!]
\includegraphics*[height=.1\textheight]{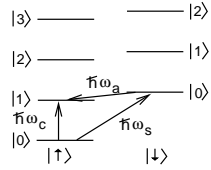}
\caption{The energy levels of the quantum cyclotron. The 
energy difference between the spin-flip transition and the
cyclotron excitation arises because $g \neq 2$ ($a \neq 0$). }
\label{fg:qc}
\end{figure}

Recently Gabrielse has developed a cooled cylindrical trap which
he calls a quantum cyclotron.\cite{gaboeo}  The trap is cooled
to mK temperatures, so that thermal excitation of the quantum cyclotron
levels is very improbable, and the axial motion is
reduced.  The quantum levels of this system are
shown in Fig. \ref{fg:qc}, where the three frequencies now become
quantum transitions. The spin flip transition is slightly different
in energy from the cyclotron energy since $a \neq 0$.
The anomaly is determined by the ratio of two frequencies
\begin{equation}
{\omega_a \over \omega_c} = { g-2 \over 2}=a; 
\quad {\rm where}\quad \omega_a \simeq 170\ {\rm MHz};
\quad \omega_{s,c} \simeq 160 \ {\rm GHz}
\end{equation}
without the need for fundamental constants.
The goal of is an order of magnitude improvement on
the precision of $a_e$, or $\sim \pm 0.3$ ppb.

\begin{figure}
\includegraphics*[width=1.9in,angle=270]{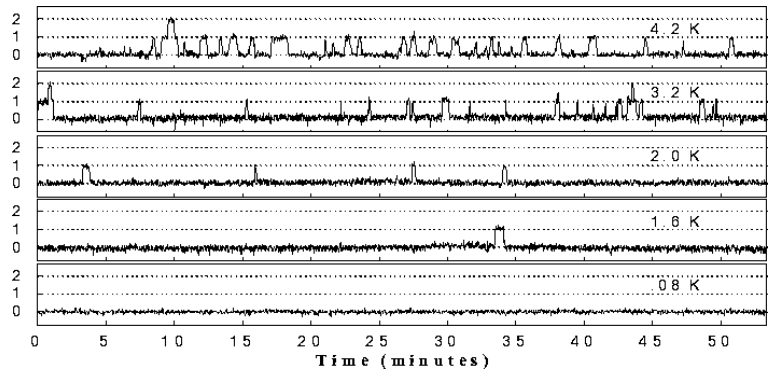}
\caption{Thermal excitation of the levels of the quantum cyclotron.
As the trap is cooled, the electron spends more time in the ground state.
At 0.08 K the electron can spend on the order of two hours without
a thermal excitation (courtesy of G. Gabrielse).}
\label{fg:qctherm}
\end{figure}

In the cylindrical trap, spontaneous emission is suppressed by 
two orders of magnitude.  Thus one can stimulate these quantum transitions
and measure the two frequencies.
The benefits of cooling the trap can be seen in Fig. \ref{fg:qctherm}
where transitions to the first and second energy states are observed
until the temperature is lowered to 0.08 K.
The trap is built and working,
and the first results can be expected in late 2003.

\section{EDM Searches, Especially for the Muon}

While the MDM has a substantial standard model value, the predicted EDMs
for the leptons are unmeasurably small and lie orders of magnitude below
the present experimental limits given in Table \ref{tb:edm}.
Thus the presence of an EDM at a measurable level would signify physics 
beyond the standard model.  Since the presently known {\sl CP} violation
is inadequate to describe the baryon asymmetry in the universe, additional
sources of {\sl CP} violation should be present.
SUSY models do predict an EDM.\cite{bdm}  For a new physics contribution
to $a_{\mu}$ of $3 \times 10^{-9}$ (of the order which might have been
seen in E821 before the CDM2 normalization error was found).
\begin{equation}
d_{\mu}^{\rm NP} \simeq 3 \times 10^{-22}\left({a_{\mu}^{ \rm NP} \over
3 \times 10^{-9}} \right)
\tan \phi_{CP} \ \ e{\rm -cm}
\end{equation}
where  $\phi_{CP}$ is a {\sl CP} violating phase.

\begin{table}[h!]
\centering
\begin{tabular}{ccc} \hline
      \tablehead{1}{c}{b}{ Particle} 
    & \tablehead{1}{c}{b}{{ Present EDM Limit}\\ (e-cm)} 
    & \tablehead{1}{c}{b}{ Standard Model \\{ Value} (e-cm)}\\
\hline
n\cite{nedm} &{$6.3 \times 10^{-26}$ } & {$10^{-31}$ }  \\
\hline
 $e^-$\cite{eedm}  & {$\sim 1.6 \times 10^{-27 }$} & {$10^{-38}$ } \\
\hline
 {$\mu$}\cite{cern3} &{$<10^{-18}$ } (CERN) & {$10^{-35}$ }\\
 & $\sim10^{-19}$ (E821)\tablenote{Estimated limit, work in progress.} \  & \\
 & {$\sim10^{-24}$ }\  \tablenote{Proposed new dedicated experiment.\cite{loi}}
 \ \ \ \ \ \  \ \ \  \ \ \  \\
\hline
\end{tabular}
\caption{Measured limits on electric dipole moments, and their standard
model values}
\label{tb:edm}
\end{table}

\begin{figure}[h!]
\includegraphics*[height=.17\textheight]{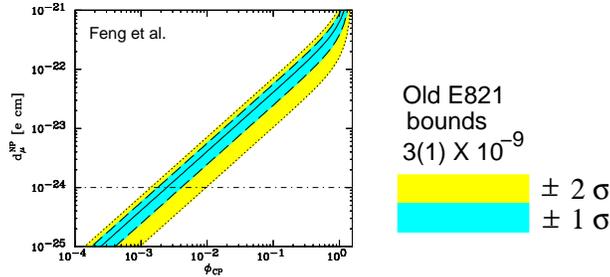}
\caption{The range of  $\phi_{CP} $ available to a dedicated 
muon EDM experiment.\cite{fm}  
The two bands show the one and two standard-deviation
ranges if $a_{\mu}$ differs from the standard model 
value by  $(3\pm 1)\times 10^{-9}$.}
\label{fg:phicp}
\end{figure}

A new experiment to search for a permanent EDM of the muon 
with a design sensitivity of $10^{-24}$ $e$-cm is now being
planned for J-PARC.\cite{loi}  This sensitivity lies well within 
values predicted by SUSY models.\cite{bdm}  Feng, et al.,\cite{fm} have 
calculated the range of $\phi_{CP} $ available to such an experiment,
which is shown in Fig. \ref{fg:phicp}.

With an EDM present, the spin precession relative to the momentum is given 
by
\begin{equation}
\vec \omega  = 
 -{e \over m} 
\left[ a_{\mu} \vec B -
\left( a_{\mu}- {1 \over \gamma^2 - 1} \right) {{\vec \beta \times \vec E }\over c }
\right] 
+
{e \over m}\left[ {\eta \over 2} \left( {\vec E \over c} +
\vec \beta \times \vec B \right) \right] 
\label{eq:omegawedm}
\end{equation}
where
$
d_{\mu} = {\eta\over 2} ({e \hbar \over 2 m c }) \simeq \eta \times 4.7\times 
10^{-14} \ \ e{\rm  - cm}
$
and 
$a_{\mu} = ({g-2 \over 2})$. For reasonable values of 
$\beta$, the motional electric field 
$\vec \beta \times \vec B$ is much larger than electric fields that can be
obtained in the laboratory, and the two vector frequencies are orthogonal
to each other. The EDM has two effects on the precession:
the magnitude of the observed frequency is increased, and the
precession plane is tipped relative to the magnetic field.

E821 was operated at the magic $\gamma$ so that the focusing 
electric field did not cause a spin precession.  The EDM signal
in E821 is very difficult to observe, since the tipping of the 
precession plane is very small ($\leq 5$ mrad).  The dedicated 
experiment will be operated at 500 MeV/c, off of the magic $\gamma$, 
and will use a radial electric field to stop the $(g-2)$ precession.
Thus the EDM would cause a steady build-up of the spin out of the
 plane with time.  Detectors would be placed above and below the storage 
region, and a time-dependent up-down asymmetry would be the signal of
an EDM.

\section{Summary and conclusions}

Measurements of the muon and electron anomalies played an important
role in our understanding of sub-atomic physics in the 20th century.
The electron anomaly was tied closely to the development of QED.
The subsequent measurement of the muon anomaly showed that the muon
was indeed a ``heavy electron'' which obeyed
QED.\cite{cern3}  With the sub-ppm accuracy now available for the 
muon anomaly,\cite{bennett} 
there may be indications that new physics is beginning to
appear.  Marciano\cite{mar} has pointed out that using a few very well
measured standard model parameters, rather than a global fit to all
electroweak measurements, predicts a Higgs mass which is much smaller
than the present experimental limit from LEP. 
If one argues that any discrepancy
between the standard model value of $a_{\mu}$ and the experimental
one is an indication 
that the hadronic contribution has been underestimated, and uses
this discrepancy to ``determine the hadronic contribution'',
the Higgs mass limit gets even smaller.  Marciano concludes that
``hints of `New Physics' may be starting to appear in
quantum loop effects.''

The non-observation of an electron EDM 
is becoming an issue for supersymmetry, just as the non-observation
of a neutron EDM implies such a mysteriously (some would say
un-naturally) small 
$\theta$-parameter for QCD. The search for EDMs will continue, and
if one is observed, the motivation for further searches in other systems
will be even stronger.  The muon presents a unique opportunity
to observe an EDM in a second-generation particle, where the {\sl CP}
phase might be different from the first generation, or the scaling
with mass might be quadratic rather than linear.

It is clear that the study of lepton moments (and neutron
EDM searches) will continue to be 
a topic of great importance in the first part of the 21st century.
Both the theoretical and experimental situations are evolving.
Stay tuned for further developments.

\begin{theacknowledgments}
I wish to thank my colleagues on the muon \g2 experiment,
 as well as 
M. Davier, E. de Rafael, G. Gabrielse, W. Marciano, and B. Odum  for
helpful discussions. 
Special thanks go to R. Carey and D. Hertzog for their helpful comments
on this manuscript.
\end{theacknowledgments}





\end{document}